\documentclass[aps,twocolumn,showpacs,superscriptaddress,groupedaddress]{revtex4}
\usepackage{amssymb, amsmath}   

\usepackage{graphicx}
\usepackage{dcolumn}
\usepackage{bm}

\usepackage{xcolor}
\usepackage{soul}
\newcommand{\BF}[1]{{\color{black} #1}}

\newcommand{\RNum}[1]{\uppercase\expandafter{\romannumeral #1\relax}}
\definecolor{Dgreen}{rgb}{0.15, 0.45, 0.15}

\newcommand{\cm}{cm$^{-1}$}
\definecolor{Dsur}{rgb}{0.85, 0.45, 0.05}

\begin{document}


\title{\BF{Coherent Dynamics of the Off-Diagonal Spin-Boson Model in the Ultra-Strong Coupling Regime}}

\author{Nirmalendu Acharyya}
  \author{Martin Richter}
   \altaffiliation[Current Address:]{ Institut f\"ur Physikalische Chemie, Friedrich-Schiller-Universit\"at Jena, D-07743 Jena, Germany}
\author{Benjamin P. Fingerhut}%
 \email{fingerhut@mbi-berlin.de}
\affiliation{%
 Max-Born-Institut f\"ur Nichtlineare Optik und Kurzzeitspektroskopie, D-12489 Berlin, Germany.
}%

\date{\today}

\begin{abstract}
Quantum mechanics describes the unitary time evolution of closed systems. In practice, every quantum system interacts with the environment leading to an irreversible loss of coherence.
The Spin-Boson model (SBM) is central to the understanding of the fundamental  process of decoherence of a two-state quantum system interacting with a bosonic heat bath but 
\BF{the nature of transient dynamics in the presence of hybrid diagonal and off-diagonal  system-bath interactions remains largely unexplored.}
Here, we investigate how the hybrid system-bath interactions of an Ohmic environment induce localization  in the bias-free SBM. 
%
\BF{For strong coupling to the environment, localization is strongly affected 
by a dynamically generated
bias via the renormalization of the tunneling amplitude.}
\BF{We find that counteractive effects of Hamiltonian parameters on non-exponential short-time dynamics and long-time population equilibration can lead to a separation of timescales and}
\BF{n}on-equilibrium quantum coherent dynamics 
that can persist even for ultra-strong system-bath interaction. The findings offer novel opportunities to exploit coherence as a resource in quantum devices operating in the ultra-strong coupling regime.
\end{abstract}

\pacs{Valid PACS appear here}
\maketitle

\section{Introduction}
The Spin-Boson model (SBM) is paradigmatic for describing  most important physical and chemical processes, like proton transfer in liquid phase \cite{Cukier:JCP:1989}, electron transfer and exciton transport in biological environments \cite{Xu:ChemPhys:1994,Thorwart:CPL:2009} and  tunneling in macroscopic two-state systems \cite{Han:PhysRevLett:1991}.
Despite its conceptual simplicity,  distinct regimes of  system-bath interaction strength $\alpha$ characterize its complex dynamics and ground state properties. 
For Ohmic dissipation and low temperature ($T\to 0$),
coherent dynamics is observed at weak system-environment interaction strength ($\alpha \lesssim \alpha_* \approx 0.5$) \cite{Leggett:1987},
mediated by tunneling amplitude $\Delta_0$. Incoherent decay arises for strong coupling to the environment ($\alpha_* \lesssim\alpha \lesssim \alpha_c \approx 1+ O(\Delta_0/\omega_c)$)\cite{Strathearn:2018}. In both, the weak and strong coupling regime, the equilibrium ground state is delocalized. Ultra-strong coupling to the environment is realized for  $\alpha >\alpha_c$. In this regime, the  tunneling amplitude renormalizes to zero  ($\Delta_r \to 0$) and induces a freezing of the dynamics at  initial configuration \cite{Anderson:1970, Magazzu:NatCom:2018}. As a consequence,  the ground state  in the ultra-strong coupling regime is localized, yielding the delocalized-to-localized BKT phase transition at $\alpha =\alpha_c$~\cite{Bray:1982,Chakravarty:1982,Leggett:1987}.
Recent  realizations of the ultra-strong coupling regime of light-matter interaction \cite{Wallraff:2004,Frisk-Kockum:NatureReviewsPhysics:2019,FornDiaz:RevModPhys:2019}  have spurred renewed interest in the SBM \cite{Forn-Diaz:NatPhys:2016,Magazzu:NatCom:2018,Frisk-Kockum:NatureReviewsPhysics:2019}.

\BF{The environment affects the two-state quantum system of the SBM via the bath induced fluctuations of localized states (i.e. eigenstates of $\sigma_z$, cf. eq.~1), inducing population relaxation and dephasing~ \cite{Weiss:book}.}
\BF{The influence of non-diagonal system-bath interactions  on the tunneling amplitude $\Delta_0$ 
is less well understood.}
\BF{Early work by Laird, Budimir, and Skinner investigated a model of two nondegenerate quantum states coupled linearly and off-diagonally to a bath~\cite{Laird:JCP:1991}.}
\BF{Their findings of strictly non-zero population excitation rates were later confirmed by Reichman and Silbey~\cite{Reichman:JCP:1996}.}
%
%
\BF{Ground state properties of the off-diagonal SBM have been investigated~\cite{Guarnieri:2018,Zhao:2014,Zhou:PhysRev:2015}.} 
\BF{Particular importance of diagonal and off-diagonal contributions  to the system-bath interaction was demonstrated recently via the finding 
of persistent steady state coherences 
in absence of a bare tunneling amplitude $\Delta_0$~\cite{Guarnieri:2018}.}
%

The nature of the transient dynamics in presence of hybrid diagonal and non-diagonal system-bath interaction remains largely unexplored.
%
%
%
Here, we demonstrate
the emergence of coherent 
dynamics facilitated by hybrid system-bath interactions that can persist even for ultra-strong coupling to the environment.
We start by describing the numerical treatment of the off-diagonal SBM with hybrid system-environment interactions (Sec.~\ref{Sec:SBM}). 
Equilibrium localization is analyzed for high and low temperature in Sec.~\ref{Sec:EqLoc}, revealing
counteractive effects of Hamiltonian parameters on non-exponential short-time dynamics and long-time population equilibration.
Exploiting the counteractive control parameters, 
 a timescale separation in short and  long-time dynamics can impose 
quantum coherent dynamics  even at ultra-strong system-bath interaction (Sec.~\ref{sec:Coh}).
Oscillation frequency and dephasing behavior are rationalized with help of the off-diagonal primary reaction coordinate (PRC) model that facilitates analytical access in the limit of ultra-slow dissipation
 
\section{Off-Diagonal Spin Boson Model}
\label{Sec:SBM}
We consider the symmetric SBM where a degenerate two-state (spin) system interacts bi-linearly with a harmonic reservoir via diagonal and off-diagonal interactions
\begin{eqnarray}
\begin{array}{lll}
H =\frac{\Delta_0}{2} \sigma_x  +  \sum_i \omega_i a_i^\dagger a_i+\frac{ b I+\cos \varphi \, \sigma_z + \sin \varphi \, \sigma_x}{2}
 \sum_i c_i X_i .
\end{array}
\label{Hamiltonian1}
\end{eqnarray}
 Here, $\sigma_\alpha$ with $\alpha=x,z$ denote Pauli matrices 
and $a_i (a_i^\dagger)$ are annihilation (creation) operators of bosonic modes with frequencies $\omega_i$ and $X_i = a_i +a_i^\dagger$.
 The mixing angle $\varphi$ interpolates between pure diagonal ($\varphi=0$) and pure off-diagonal ($\varphi=\pi/2$) coupling to the environment. 
 The  term  $b I \sum_i c_i (a_i +a_i^\dagger)$ shifts the origin of bath oscillators and controls their equilibrium displacement.
 \BF{It is well understood that for diagonal coupling to the environment ($\varphi=0$) and ergodic system-environment dynamics initial preparation effects via parameter $b$ are insignificant. In absence of initial system-bath correlations,
  the system relaxes to thermal equilibrium  independent of the initially prepared state~\cite{Weiss:book}}.
As we will show, 
\BF{the} term  \BF{$b I \sum_i c_i (a_i +a_i^\dagger)$} takes a crucial role in affecting non-equilibrium dynamics for $\varphi\neq0$.

\begin{figure*}
\centering \vspace*{-0.5cm}
\includegraphics[width=0.78\textwidth]{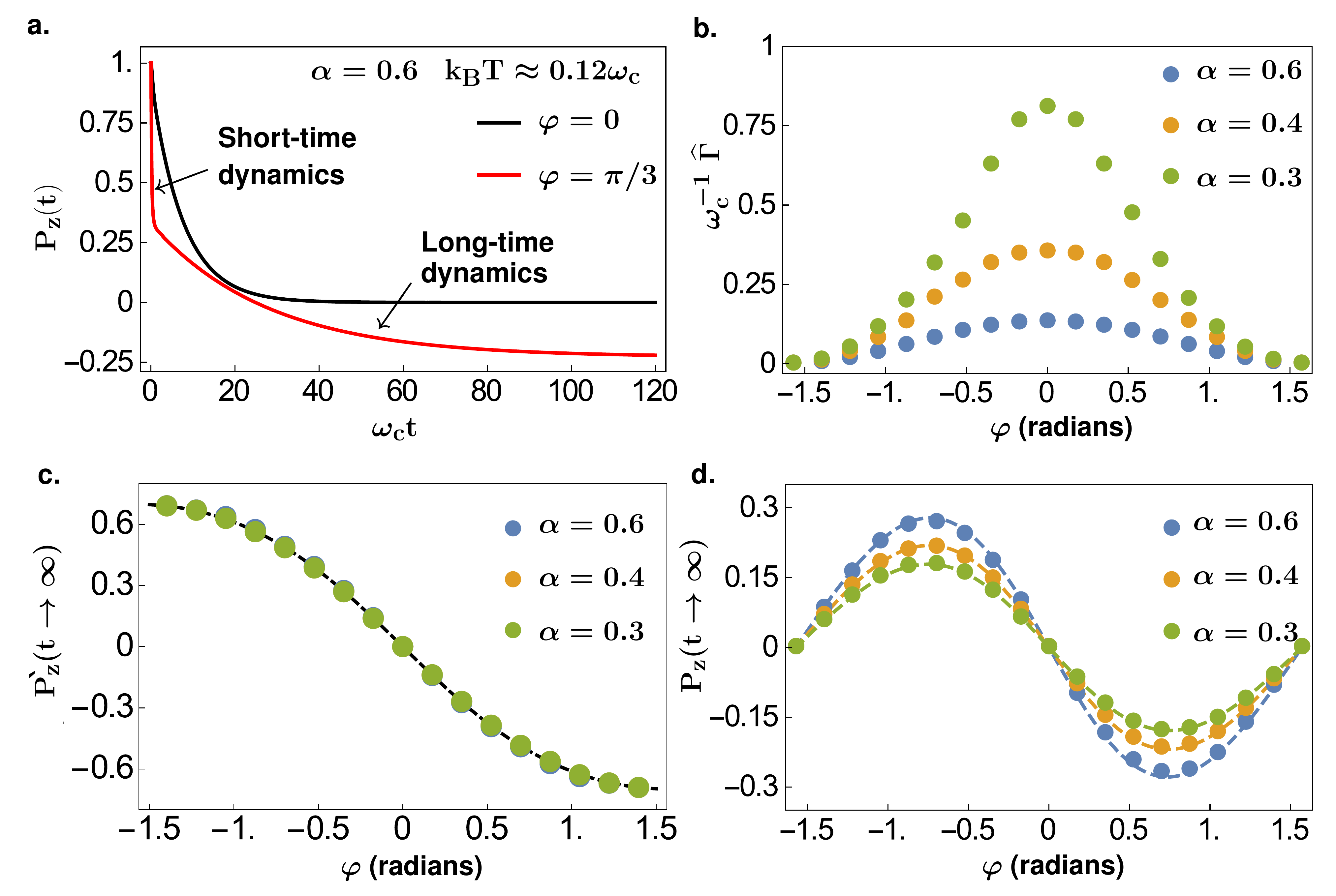}
\vspace*{-0.25cm}
\caption{
\textbf{a.} 
Impact of \BF{hybrid} 
system-bath interaction  
on  short-time and long-time dynamics\BF{: population difference  $P_z(t)=  Tr_B[\sigma_z \widehat{\rho}(t)]$} 
\BF{for diagonal ($\varphi = 0$) and hybrid diagonal and off-diagonal coupling  ($\varphi \neq 0$)}
in the overdamped regime  ($\alpha > \alpha_\ast (T)$\BF{, see SI for 
non-equilibrium dynamics at weak system-bath interaction and  $\varphi = 0$ ($\alpha\lesssim \alpha_*(T) \approx \Delta_r/(\pi k_B T)$}).
%
\BF{\textbf{b.}} \BF{Population relaxation} 
rate $\widehat{\Gamma}$ for varying system-bath interaction strengths $\alpha$, $\widehat{\Gamma}$ is obtained via \BF{an} exponential fit of \BF{the} late-time dynamics \BF{(see SI)}.
%
%
\BF{\textbf{c.}} 
Population difference  $P'_z(t\to\infty)$ in intermediate, transformed basis (cf. SI, eq.~S.1-S.3) as function of mixing angle $\varphi$ and for various values of $\alpha$ (dots).  The dashed line 
gives the analytical results with effective bias $\widehat{\epsilon}=\Delta_0 \sin\varphi$ due to basis rotation. Note that numerical values  for varying $\alpha$ almost perfectly coincide. 
%
%
\BF{\textbf{d.}} 
$P_z(t\to\infty)$  as function of mixing angle $\varphi$ and for various values of $\alpha$, dots and dashed lines 
represent numerical data and analytical results
(eq.\ref{eqn_AB} with $F=(\omega_c \beta)^{-\alpha}$), respectively.
Simulations are performed at $k_BT \approx 0.12\omega_c \gtrapprox \Delta_r$ ($T=100\,K$) and 
$\Delta_0=0.2~\omega_c$, $b=0$\BF{, cf. SI, Table S1.}
}
\label{Fig1}
\end{figure*}

The environment is  characterized by the spectral density 
$J(\omega) \equiv \frac{\pi}{2} \sum_j \frac{c_i^2}{m_j\omega_j} \delta(\omega - \omega_j)$.
We consider Ohmic dissipation  with Lorentzian high frequency cut-off, $J(\omega) = 2 \pi \alpha \omega {\omega_c^2}/({\omega^2+\omega_c^2})$, where $\alpha$  characterizes the system-bath interaction strength  and the cut-off frequency $\omega_c$ is related to 
 the inverse of Drude memory time $\tau_D=1/\omega_c$  \cite{Weiss:book,Magazzu:NatCom:2018}.

 Observables are determined by the 
 reduced density matrix $\widetilde{\rho}(t) = Tr_B \Big[ e^{-i Ht} \rho(0) e^{i Ht}\Big]$. Employing factorized initial conditions 
 and assuming the bath in thermal equilibrium, 
 $\widetilde{\rho}(t)$ was evaluated numerically for initial condition $\widetilde{\rho}(0)=|+\rangle \langle +|$ ($|\pm \rangle$ denote eigenstates of $\sigma_z$)
  with the non-perturbative quasi-adiabatic propagator path integral method \cite{Makri:JCP:1995,Sim:CPC:1997,Sim:JCP:2001}\BF{. Mask} 
  assisted coarse graining of influence coefficients  (MACGIC-QUAPI)~\cite{Richter:2017}   facilitates access to long-time non-Markovian system-bath correlations.
 %
 The algorithm exploits a finite memory time $\tau_M \propto \tau_D$ characterizing non-Markovian memory time scale and  uses a coarse grained representation of the influence functional (represented by mask of size $k_{eff}$) for computational efficiency.  By decreasing the size of the Trotter  time-step $\Delta t$ and increasing memory time $\tau_M = \Delta k_{max} \Delta t$ convergence to  numerically exact results is obtained via an increase in the  number of coarse grained quadrature points ($k_{eff} \rightarrow \Delta k_{max}$)
 and a decrease in filter threshold ($\theta \rightarrow 0$),
details are given in Refs.~\citenum{Richter:2017,Richter:FaradayDisc:2019}\BF{~(see SI for convergence of numerical simulations)}.


\section{Equilibrium localization with Off-Diagonal System-Environment Interactions}
\label{Sec:EqLoc}

%
Fig.1\BF{a}
demonstrates the impact of \BF{hybrid} 
system-environment interaction on non-equilibrium dynamics 
 in the strong  coupling regime \BF{at high temperature} 
 ($\alpha\gtrsim\alpha_*(T)$\BF{$\approx \Delta_r/(\pi k_B T)$}, \BF{$k_BT \gtrapprox \Delta_r$ with 
renormalized tunneling amplitude $\Delta_r$~\cite{Leggett:1987}}). Two distinct dynamical regimes can be identified: 
\BF{ultra-fast universal decoherence is reflected in non-exponential decay~\cite{Makarov:CPL:1994,Braun:PhysRevLett:2001,Tuorila:PhysRevResearch:2019}} and is followed by relatively slower long-time equilibration dynamics.
\BF{The amplitude of ultra-fast non-exponential decay shows pronounced sensitivity to the non-diagonal interaction ($\varphi\neq0$).}
%
\BF{It is apparent from Fig.1a that the}  off-diagonal system-environment interaction ($\varphi\neq 0$) \BF{also} affects the equilibrium ground state \BF{$P_z(t \to \infty)$}, i.e.,
the system equilibrates to a localized state ($P_z(t \to \infty) \neq 0$) even  in absence of a bare bias 
(\BF{cf.~}eq.\ref{Hamiltonian1}). 
In principle, two distinct effects  determine equilibrium localization: (i) an effective bias $\widehat{\epsilon}=\Delta_0 \sin\varphi$ is generated via a rotation of basis due to non-diagonal system environment interaction \cite{Romero-Rochin:PhysicaA:1989,Lu:JCP:2013} (see also SI, eq.~S1-S3);
(ii) renormalization of the tunneling amplitude $\Delta_0$ due to the strong interaction with the environment affects  the dressed Hamiltonian ~\cite{Weiss:book}. 
\emph{A priori}, the relevance and interdependence of both effects is unclear and has not been explored numerically.
%
%

We find that $P_z(t \to \infty)$ shows a strong dependence on $\varphi$ and increases with  coupling strength $\alpha$ (Fig.1\BF{d}).
%
\BF{Localization for $\varphi\neq 0$ is determined by the amplitude of ultra-fast non-exponential decay and  long-time exponential dynamics. The latter is characterized by the population relaxation rate $\widehat{\Gamma}$ which decreases with increasing $\alpha$ ~\cite{Weiss:book} and additionally 
 is a non-monotonic function of $\varphi$ that decreases as $\varphi \to \pi/2$  (Fig.1b).}
%
%
\BF{The pure dephasing case 
is realized for $\varphi = \pi/2$ where the tunneling amplitude, and consequently  $\widehat{\Gamma}$ vanishes.}


Equilibrium localization was analyzed 
by \BF{extending}
a generalized master equation approach \cite{Leggett:1987, Grifoni:1999} to derive approximate solutions of exponential long-time dynamics \BF{in presence of diagonal and  non-diagonal system-environment interaction ($\varphi\neq0$)}.
We 
\BF{therefore} unitarily transform the Hamiltonian (eq.\ref{Hamiltonian1}) $H \to \widehat{H} = U^\dagger H U$ with $U=\exp(-i \frac{\varphi}{2} \sigma_y)$, where 
U diagonalizes the system-bath interaction \cite{Lu:JCP:2013}.
$\widehat{H}$ takes the form of the ordinary SBM with transformed initial conditions and transformed 
parameters $\widehat{\epsilon}=\Delta_0 \sin\varphi$ and $\widehat{\Delta}=\Delta_0 \cos \varphi$. 
\BF{Note that 
$U$ is defined in space of system-bath interaction and does not necessarily diagonalize the system part of eq.~1 as both contributions to the total Hamiltonian do not commute.}
Approximate solutions for long-time dynamics were obtained in transformed basis, followed by reverse transformation. In the overdamped 
regime  this yields exponential dynamics
 \begin{equation}
 P_z(t) \approx (A -B\tanh (-\beta E/2))e^{ -\widehat{\Gamma} t} +B\tanh (-\beta E/2)
 \label{eq:Pz}
 \end{equation}
with amplitudes
 \begin{equation}
\begin{array}{ll}
 &A= \frac{(1-F)^2 \sin^2 \varphi \cos^2\varphi}{\sin^2 \varphi+ F^2 \cos^2 \varphi},  \quad B = \frac{(1-F)\sin \varphi \, \cos \varphi}{\sqrt{\sin^2 \varphi+F^2 \cos^2 \varphi}} 
\end{array}
\label{eqn_AB}
\end{equation}
where $E \equiv \sqrt{\widehat{\epsilon}^2 + \widehat{\Delta}_r^2}$,  $\beta \equiv 1/ (k_BT)$ and $F \equiv \widehat{\Delta}_r/\Delta_0$ (see SI for derivation).

\BF{From eq.~\ref{eq:Pz}-\ref{eqn_AB} we find that equilibrium localization amplitude $P_z (t \to \infty)=B \tanh (-\beta E/2)$ crucially depends on the renormalization factor $F$. In the weak coupling regime the renormalization effect is weak  ($F \approx 1$) and $B$ is vanishingly small. Thus, localization in absence of  a bare bias arises due to a dynamically generated bias 
via the renormalized tunneling amplitude $\widehat{\Delta}_r$ at strong coupling to the environment. We note that the dynamically generated bias is distinct from the bias $\widehat{\epsilon}$ due to rotation of bare Hamiltonian parameters. This becomes evident from the fact that localization does not maximize near  $\varphi=\pi/2$, instead at high temperature we find that   $P_z(t \to \infty)$ 
maximizes near $\varphi=\pi/4$ (Fig.~1d).
For comparison, Fig.1c show the population difference  $P'_z(t\to\infty)$ in intermediate, transfomred basis with the effective bias $\widehat{\epsilon}$ due to rotation of bare Hamiltonian where  localization maximizes for $\varphi=\pi/2$~\cite{Weiss:book,Romero-Rochin:PhysicaA:1989}. 
%
}
%
\BF{Excellent agreement is found for $P_z (t \to \infty)$  obtained 
from numerical  simulations and as predicted by eq.~\ref{eqn_AB}  where the renormalization effect via $F$ is taken into account
(Fig.1d, see also SI, Fig.S3 for $P_x (t \to \infty)$ dependence on $\varphi$).}
At high temperature and strong coupling, 
$\widehat{\Delta}_r =\widehat{\Delta}(\omega_c \beta)^{-\alpha}$ \cite{Costi:1999,Ruakola:PhysRevB:2011}
which yields a non-zero decay rate $\widehat{\Gamma} \propto \Delta_0^2 \cos^2 \varphi  (\omega_c \beta)^{-2\alpha}$ and further confirms the numerically observed behavior 
(Fig.1\BF{b}).
In the off-diagonal case  \BF{at low temperature} ($\varphi \neq 0$, \BF{$k_BT \rightarrow 0$}), 
\BF{the} finite \BF{population relaxation} 
rate $\widehat{\Gamma}$ prevails  at weak and strong coupling due to \BF{a} non-vanishing renormalized tunneling amplitude
$\widehat{\Delta}_r = \widehat{\Delta} (\widehat{\Delta}/\omega_c)^{\alpha/(1-\alpha)}$ \cite{Leggett:1987,Costi:1999}.
\BF{Nevertheless, eqs.~\ref{eq:Pz}-\ref{eqn_AB} predict localization for $\varphi \neq 0$ ($P_z(t \to  \infty)\approx B  \tanh (-\beta E/2) \neq 0$ with $\widehat{\Delta}_r = \widehat{\Delta} (\widehat{\Delta}/\omega_c)^{\alpha/(1-\alpha)}$) that was confirmed numerically (Fig.~2b).}
\BF{Such findings contrast with the localization behavior induced by the effective bias $\widehat{\epsilon}=\Delta_0 \sin\varphi$  generated via basis rotation (Fig.~2a) and the 
bias-free SBM ($\varphi=0$) subject to Ohmic dissipation  that 
  shows delocalization in weak and strong coupling regimes ($P_z(t \to \infty)=0$ for $\alpha \lesssim \alpha_* \approx 0.5$ and $\alpha_* \lesssim\alpha \lesssim \alpha_c \approx 1+ O(\Delta_0/\omega_c)$~\cite{Strathearn:2018}).
} 
%

\begin{figure*}\centering\vspace*{-0.25cm}
\includegraphics[width=0.75 \textwidth]{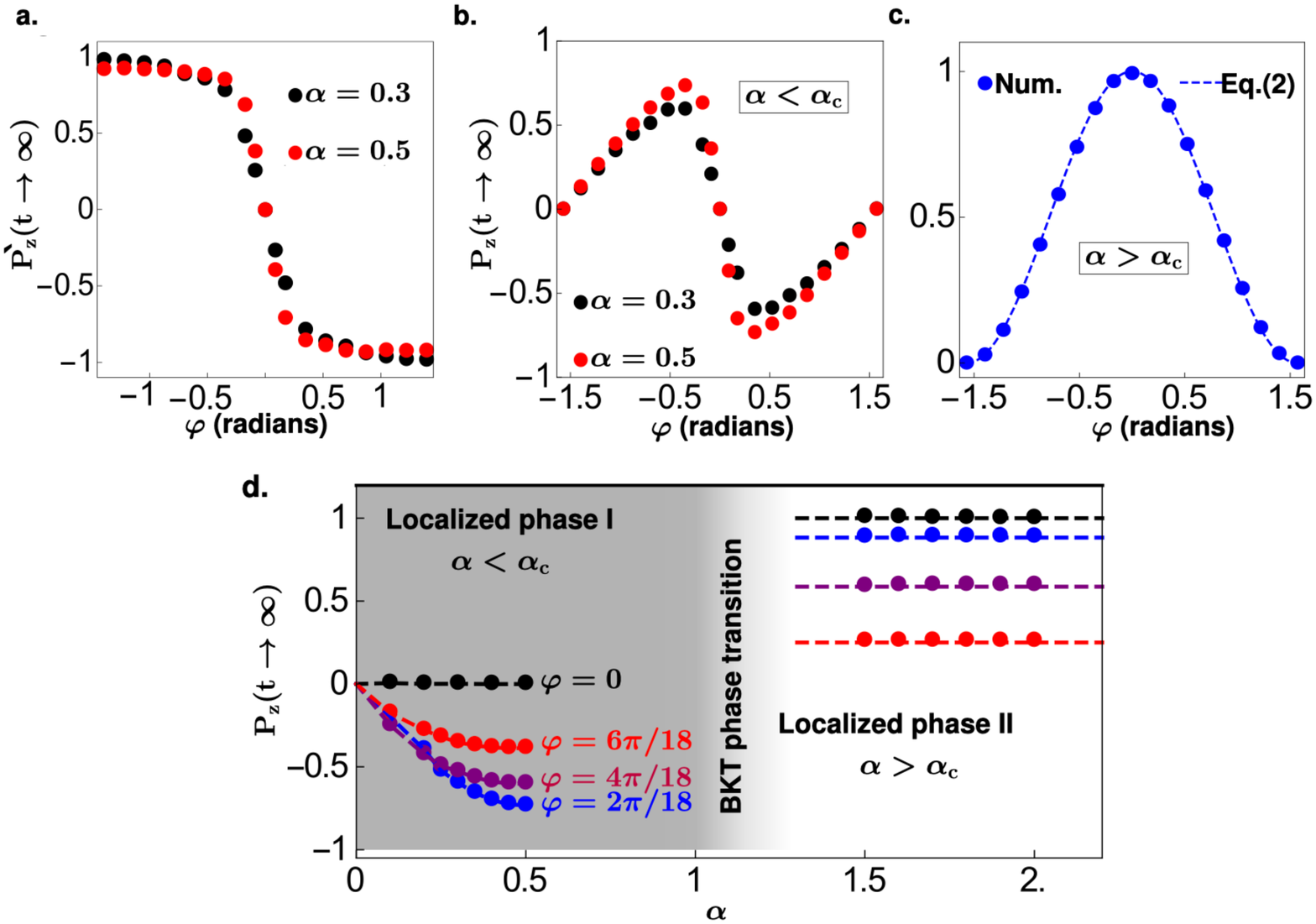}
\vspace*{-0.25cm}
\caption{
 (\textbf{a.})
Population difference  $P'_z(t\to\infty)$ in intermediate, transformed basis (cf. SI, eq.~S.1-S.3) as function of mixing angle $\varphi$
at  low $T\to 0$
 for 
 $\alpha < \alpha_c$.
$P_z(t \to \infty)$  as a function of $\varphi$ 
at  low $T\to 0$
 for 
 $\alpha < \alpha_c$ (\textbf{b.}) and $\alpha > \alpha_c$ (\textbf{c.}).
Dots and dashed line mark numerical and analytical results (eq.\ref{eqn_AB} with  $F=0$), respectively. 
\textbf{d.} Sketch of the 
\BF{localized-to-localized} phase transition in the off-diagonal SBM.
Phase \RNum{1} ($\alpha<\alpha_c$) shows finite values of $P_z(t \to \infty)$ 
 for $\varphi \neq 0$ and dependence on $\alpha$, $\varphi$, and $\Delta_0/\omega_c$. 
In  
phase \RNum{2} ($\alpha>\alpha_c$), $P_z(t \to \infty)$ is determined by $\varphi$ but independent of $\alpha$ and $\Delta_r$.  
In numerical simulations $\Delta_0/\omega_c=0.1 $, $b = 0$ and $k_BT \approx 0.0003 \omega_c$ ($T \approx 0.05 K$), \BF{cf. SI, Table S2-S5.}
%
%
}
\label{Fig2}
\end{figure*}

\BF{A}t ultra-strong coupling and 
low temperature 
the tunneling amplitude
renormalizes to zero  \BF{for $\varphi = 0$} ($\Delta_r \to 0$\BF{, $\alpha \gtrsim \alpha_c \approx 1+ O(\Delta_0/\omega_c)$, $k_BT \rightarrow 0$}) \cite{Anderson:1970, Magazzu:NatCom:2018} which leads to freezing of the population at the initial configuration and formation of a localized phase.\cite{Strathearn:2018}
For $\varphi \neq 0$, $\Delta_r \to 0$ behavior 
is preserved ($F=0$, eq.\ref{eqn_AB}) beyond the same value of $\alpha _c$ 
\footnote{The RG flow equations in transformed basis for $\widehat{\Delta}$  with $\varphi\neq 0$ are identical to that of ${\Delta}_0$ with $\varphi= 0$ \cite{Anderson:1970,Florens:2010}: 
$\frac{d (\widehat{\Delta}/\omega_c)}{d \ell} \approx (1-\alpha) \widehat{\Delta}/\omega_c$ and $\frac{d \alpha}{d \ell} \approx -\alpha (\widehat{\Delta}/\omega_c)^2$.
This ensures that the fixed point $(\widehat{\Delta} \approx 0, \alpha \approx1)$ for $\varphi\neq 0$ and $\varphi= 0$ remains identical. 
Consequently, the BKT phase transition for any $\varphi$ emerges at $\alpha = \alpha_c = 1+ O (\Delta_0/\omega_c)$.}.
Accordingly, $\widehat{\Gamma} \to 0$ 
for $\alpha> \alpha_c$ and the mechanism leading to freezing of the dynamics in the SBM \cite{Bray:1982, Chakravarty:1982} is conserved  for $\varphi \neq 0$.
 Localization amplitude $P_z(t \to \infty) =A =\cos^2 \varphi $ (obtained upon inserting $\widehat{\Gamma}=0$ and $F=0$  in eq.\ref{eqn_AB}) is confirmed with excellent accuracy in numerical simulations (Fig.2c).
%

As a first important result we thus find
a BKT phase transition
for $\alpha \approx
\alpha_c$ and all values of $\varphi$\BF{, however},
  for $\varphi \neq 0$ the transition is between distinct 
  \BF{localized} phases  \RNum{1} and \RNum{2} (Fig.2d) with different
localization mechanism and ground state.
In phase \RNum{1} ($\alpha<\alpha_c$), equilibrium 
properties depend on $\Delta_0/\omega_c$, $\alpha$ and $\varphi$. 
Localization arises \BF{due to} 
a dynamically generated 
bias 
via the renormalization of $\widehat{\Delta}_r$ \BF{transcending behavior of the SBM in the weak coupling regime}. 
In  phase \RNum{2}  ($\alpha>\alpha_c$) 
$\Delta_0/\omega_c$ become irrelevant. 
Localization  sensibly depends on $\varphi$ but  is found to be independent of $\alpha$ 
\footnote{For ultra-strong coupling ($\alpha>\alpha_c$) and $\varphi=0$, localization implies a degenerate ground state~\cite{Bray:1982, Chakravarty:1982}. 
Specifically, changing the initial condition from $P_z(t=0)=1$ to $P_z(t=0)=-1$, yields that $P_z(t\to\infty)=1$ changes to $P_z(t\to\infty)=-1$. 
Such degeneracy is preserved for $\varphi\neq 0$:  for initial condition from $P_z(t=0)=\pm1$, the system equilibrates to $P_z(t\to\infty)=\pm \cos^2 \varphi$.}.
In both phases, the term $b I \sum_i c_i (a_i +a_i^\dagger)$ (eq.\ref{Hamiltonian1}) has no effect on long-time dynamics, i.e., 
$\widehat{\Gamma}$ and 
$P_z(t \to \infty)$. 

\BF{The 
population relaxation rate $\widehat{\Gamma}$ decreases with increasing $\alpha$ and $\varphi \to \pi/2$ (Fig.1b).
In contrast, we observe for any temperature an acceleration of short-time dynamics as $\alpha$  increases and $\varphi \rightarrow \pi/2$  (SI, Fig.S4).
Such counteractive effects of $\alpha$ and $\varphi$ on short- and long-time dynamics can be exploited for control of dynamics beyond the scope of 
the ordinary SBM with Ohmic dissipation.}

\begin{figure*}[ht]\vspace*{-0.45cm}
\hspace*{-0.5cm}\includegraphics[width=0.82 \textwidth]{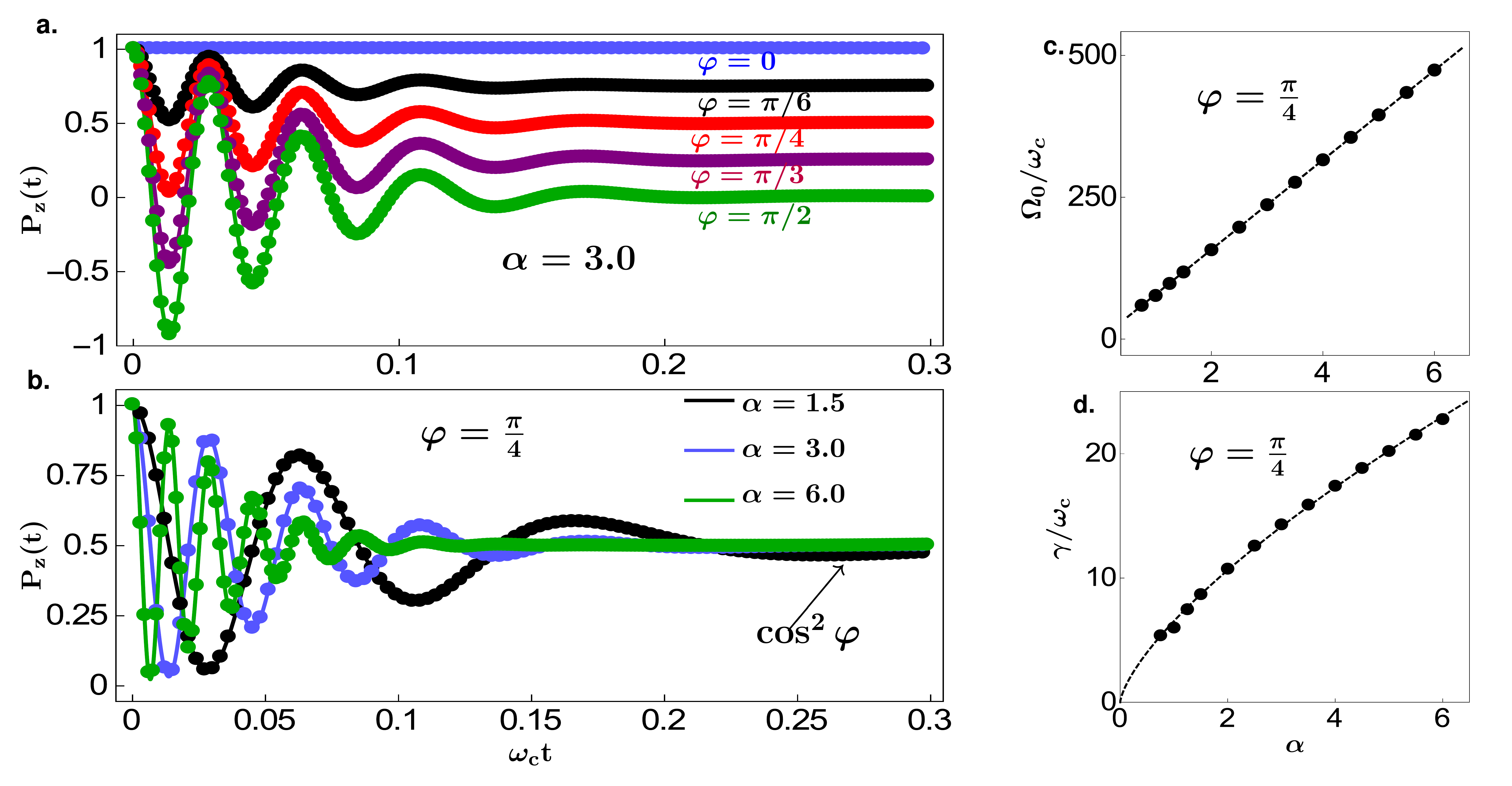}
\vspace*{-0.5cm}
\caption{
Non-equilibrium  
dynamics at low temperature in the ultra-strong coupling regime ($\alpha >  \alpha_c$)  
for various values $\varphi$ (\textbf{a.}) 
and 
 $\alpha$ (\textbf{b.}).
Solid lines represent MACGIC-QUAPI simulations and dots mark OFP simulations (see SI). 
Dependence of  oscillation frequency $\Omega_0$ (\textbf{c.}) and damping constant $\gamma$  (\textbf{d.}) on coupling strength  $\alpha$.
In all simulations $\Delta_0=0.1 \omega_c$, $b = 4$ and $T \approx 0.05 K$, 
\BF{cf. } SI, Table~S6.
}
\label{Fig3}
\end{figure*}

\section{Coherent Dynamics at Ultra-Strong System-Environment Interaction}
\label{sec:Coh}


Fig.\ref{Fig3} presents  
non-equilibrium dynamics of the off-diagonal SBM ($\varphi \neq 0$) with ultra-strong coupling 
($\alpha>\alpha_c$, $k_BT \approx 0.0003\omega_c$).
For diagonal system-bath interaction ($\varphi = 0$) the dynamics 
remains frozen in the initial configuration. In contrast, for $\varphi\neq 0$ 
nontrivial short-time dynamics precedes freezing of  dynamics for $\alpha>\alpha_c$. Interestingly, such short-time dynamics can be oscillatory for $\varphi\neq 0$ and  $b\neq 0$, despite the ultra-strong 
interaction strength (Fig.3a,b). 
%
%
At the origin of oscillatory short time dynamics is a timescale separation, i.e., particularly fast short-time dynamics and slow long-time equilibration \BF{due to counteractive effects 
of parameters $\alpha$, $\varphi$ and $b$
on system, interaction and bath part of the Hamiltonian (eq.~1). Universal decoherence \cite{Braun:PhysRevLett:2001} imposes non-exponential short-time dynamics that is independent of the system Hamiltonian but } 
strongly affected by 
\BF{the details of the system bath interaction and the bath}
(\BF{$\alpha$}, $\tan \varphi$ \BF{and $b$}).
For any temperature we observe acceleration of the \BF{short-time} dynamics as $\alpha$  increases, $\varphi$ approaches $\pi/2$ and $b$ increases (SI, Fig.S\BF{4}).
Parameter $b$ \BF{of the bath Hamiltonian} is thus a new relevant parameter for the off-diagonal SBM.
\BF{Conversely,   $\alpha$  and $\varphi$ affect the system dynamics (population relaxation and dephasing) via basis rotation and renormalization of tunneling amplitude $\Delta_0$.}
%


Observed dynamics exhibits the following characteristics: (i) the oscillation amplitude  sensibly depends on $\varphi$ (Fig.3a); (ii) the oscillation frequency increases linearly with increasing $\alpha$ (Fig.3b,c) and $b$ (SI, Fig.S\BF{6});
 (iii) short-time dynamics is independent   of $\Delta_0/\omega_c$  and insensitive to a further lowering of $T$ (SI, Fig.S\BF{5})
 which 
 \BF{reflects the} quantum coherent origin \BF{of universal decoherence for $\varphi \neq 0$}. 
 The oscillatory dynamics has thus a fundamentally different origin than the one observed in the weak coupling regime ($\alpha < \alpha_\ast$)  where oscillation frequency is determined by $\Delta_r$.

We find that the  fit-function 
\begin{equation}
P_z(t)\approx\cos^2 \varphi + \sin^2 \varphi \, \cos(\Omega_0 t + \Omega_1 t^2 + \Omega_2 t^3 ) e^{- \gamma^2 t^2}
\label{Pz_empirical}
\end{equation}
 accurately describes the dynamics.
 The $\cos^2 \varphi$ term 
 accounts for long-time behavior  ($P_z(t \rightarrow \infty)$), $\Omega_0$ describes the leading order oscillation frequency,
   $\Omega_{1,2}$~account for temporal modulation of 
$\Omega_0$
  and $\sin^2\varphi$ determines the amplitude. The latter  vanishes for $\varphi=0$
 rendering off-diagonal coupling necessary for emerging coherences.  
 A fit  of the dynamics reveals a linear  increase of $\Omega_0$   with $\alpha$ ($\Omega_0 \propto b\,\alpha\,\omega_c$, Fig.3c) which
partially resembles oscillatory behavior of sub-Ohmic environments \cite{Kast:2013}.
 Oscillations decay with $\exp[- \gamma^2 t^2]$  where damping constant $\gamma$ increases sub-linearly with $\alpha$ ($\gamma = \gamma_0 \alpha^k\, \omega_c$ with $k\approx 0.7$; Fig.3d), thus, forming the essence  of persistence of oscillations for increasing $\alpha$ (Fig.3b). 
\BF{The observed sub-exponential damping is distinct from the reported super-exponential to algebraic decay for diagonal system bath interaction~\cite{Tuorila:PhysRevResearch:2019}.}

Microscopic origin  of oscillatory short-time dynamics was rationalized via a off-diagonal 
PRC model that couples electronic states $|\pm \rangle$ to a primary coordinate $Q$ that in turn couples to a dissipative Ohmic environment \cite{Garg:JCP:1985,Lambert:NatCom:2019,Correa:JCP:2019}:
\begin{eqnarray}
&&\hspace{-0.5cm}\mathcal{H}= \frac{1}{2}\Delta_0 \sigma_x+ \widehat{\Omega}^2 \frac{\kappa}{2}(\cos \varphi  \sigma_z+ \sin \varphi \sigma_x +b )Q +\mathcal{H}_Q+\mathcal{H}_B  \nonumber \\ 
 &&\hspace{-0.5cm}\mathcal{H}_B = \sum_{j=0}^{\infty} 
 \left[\frac{p_{j}^2}{2 m_{j}} + \frac{1}{2} m_{j} \omega_{j}^2 \left(q_{j} - \frac{g_{j}Q}{m_{j } \omega_{j}^2}  \right)^2\right] 
 \label{eq:HPO}
 \end{eqnarray}
 and $\mathcal{H}_Q = \frac{1}{2}\left(P^2 + \widehat{\Omega}^2 Q^2 \right)$. Canonical transformation $Q =\sum_j v_j q_j$  allows to map
 eq.\ref{eq:HPO} onto the off-diagonal SBM (eq.\ref{Hamiltonian1}) 
 \cite{Chernyak:1996} 
 with 
 $\tan \varphi$ being a measure of non-Condon effects~\cite{Mavros:2016,Domcke:Book}.

 In the limit of ultra-slow dissipation 
 (ultra-strong coupling), the approximate time-evolution of a vibronic wavepacket in the PRC 
 (eq.\ref{eq:HPO}) confirms the leading order  oscillatory  behavior (cf. SI for analytical treatment).
The oscillation frequency $\Omega_0$ grows linearly with $b \alpha$,  the amplitude of the leading order term  being $\sin^2 \varphi$ that is damped via 
$\gamma$ which shows sub-linear dependence on $\alpha$. 

\vspace{1cm}
\section{Conclusions}
The close agreement between analytical and numerical findings suggests a picture where oscillatory dynamics, preserved in the ultra-strong coupling regime, arises via  non-Condon effects mediated by a PRC
contained in the  Ohmic spectral density. This suggests a  dominant short-time relaxation channel of the off-diagonal SBM where the initially excited state strongly interacts with  mode $Q$. 
From the perspective of the SBM this implies that at ultra-strong coupling in the short-time regime the system interacts primarily with a few bath modes.
The finite (non-Markovian)  relaxation timescale of the Ohmic spectral density constitutes a ``sluggish" environment and prevents uniform dissipation, thus,  imposing a separation of timescales. The non-equilibrium state of the PRC 
is prevented from instantaneous dissipation into the bath and the electronic system becomes  susceptible to re-excitation, resulting in oscillatory short-time dynamics that can not be rationalized via bare tunneling amplitude $\Delta_0$.
The particular high efficiency of the off-diagonal mediated dissipation channel 
has far reaching relevance for ultrafast condensed phase (molecular) relaxation where
the importance of  non-Condon effects was stressed 
in molecular vibronic photo-relaxation \cite{Tamura:JCP:2012,Christensson:2012aa,Schnedermann:NatComm:2019}
\footnote{In molecular systems  typical primary mode frequency $\widehat{\Omega} \sim 200$ \cm \cite{Hamm:PhysRevLett:2012} and damping coefficient $\gamma \sim 2500$ \cm ($\omega_c = \widehat{\Omega}^2/\gamma$ = 16 cm$^{-1}$\cite{Eckel:NJP:2009}, $\gamma \gg \widehat{\Omega}$ \cite{Chernyak:1996}),  which gives an oscillations frequency $\Omega_0 \approx 950$ \cm (eq.~\ref{Pz_empirical}) when $\alpha=1$ and $b=3$. On the other hand, when $\alpha=2.5$ and $b=3$, $\Omega_0 \approx 2370$ \cm. }.  


The prediction of quantum coherent dynamics in the ultra-strong coupling regime is readily amenable to experimental verification in various platforms \cite{Frisk-Kockum:NatureReviewsPhysics:2019}. In circuit  QED, 
the ultra-strong coupling regime with strong system-bath entanglement was demonstrated   \cite{Forn-Diaz:NatPhys:2016} and control over the longitudinal-to-transverse  coupling ratio $\tan\varphi$ is provided via Josephson and  charging energies; 
experimental manipulation of the displacement parameter $b$ 
can be achieved via the gate charge \cite{Blais:2004}.
 SQUID devices \cite{Han:PhysRevLett:1991} provide access to variations  of  $\Delta_0$  
 and thus control over the amplitude of off-diagonal induced steady state coherences and localization in the weak and strong coupling regime.

\section*{ACKNOWLEDGMENTS}
This research has received funding from the European Research Council (ERC) under the European Union's Horizon 2020 research and innovation program (grant agreement No. 802817). B.P.F. acknowledges support by the DFG within the Emmy-Noether Program (Grant No. FI 2034/1-1).

\bibliographystyle{apsrev}

\begin{thebibliography}{47}
\expandafter\ifx\csname natexlab\endcsname\relax\def\natexlab#1{#1}\fi
\expandafter\ifx\csname bibnamefont\endcsname\relax
  \def\bibnamefont#1{#1}\fi
\expandafter\ifx\csname bibfnamefont\endcsname\relax
  \def\bibfnamefont#1{#1}\fi
\expandafter\ifx\csname citenamefont\endcsname\relax
  \def\citenamefont#1{#1}\fi
\expandafter\ifx\csname url\endcsname\relax
  \def\url#1{\texttt{#1}}\fi
\expandafter\ifx\csname urlprefix\endcsname\relax\def\urlprefix{URL }\fi
\providecommand{\bibinfo}[2]{#2}
\providecommand{\eprint}[2][]{\url{#2}}

\bibitem[{\citenamefont{Cukier and Morillo}(1989)}]{Cukier:JCP:1989}
\bibinfo{author}{\bibfnamefont{R.~I.} \bibnamefont{Cukier}} \bibnamefont{and}
  \bibinfo{author}{\bibfnamefont{M.}~\bibnamefont{Morillo}},
  \bibinfo{journal}{J. Chem. Phys.} \textbf{\bibinfo{volume}{91}},
  \bibinfo{pages}{857} (\bibinfo{year}{1989}).

\bibitem[{\citenamefont{Xu and Schulten}(1994)}]{Xu:ChemPhys:1994}
\bibinfo{author}{\bibfnamefont{D.}~\bibnamefont{Xu}} \bibnamefont{and}
  \bibinfo{author}{\bibfnamefont{K.}~\bibnamefont{Schulten}},
  \bibinfo{journal}{Chem. Phys.} \textbf{\bibinfo{volume}{182}},
  \bibinfo{pages}{91 } (\bibinfo{year}{1994}).

\bibitem[{\citenamefont{Thorwart et~al.}(2009)\citenamefont{Thorwart, Eckel,
  Reina, Nalbach, and Weiss}}]{Thorwart:CPL:2009}
\bibinfo{author}{\bibfnamefont{M.}~\bibnamefont{Thorwart}},
  \bibinfo{author}{\bibfnamefont{J.}~\bibnamefont{Eckel}},
  \bibinfo{author}{\bibfnamefont{J.}~\bibnamefont{Reina}},
  \bibinfo{author}{\bibfnamefont{P.}~\bibnamefont{Nalbach}}, \bibnamefont{and}
  \bibinfo{author}{\bibfnamefont{S.}~\bibnamefont{Weiss}},
  \bibinfo{journal}{Chem. Phys. Lett.} \textbf{\bibinfo{volume}{478}},
  \bibinfo{pages}{234 } (\bibinfo{year}{2009}).

\bibitem[{\citenamefont{Han et~al.}(1991)\citenamefont{Han, Lapointe, and
  Lukens}}]{Han:PhysRevLett:1991}
\bibinfo{author}{\bibfnamefont{S.}~\bibnamefont{Han}},
  \bibinfo{author}{\bibfnamefont{J.}~\bibnamefont{Lapointe}}, \bibnamefont{and}
  \bibinfo{author}{\bibfnamefont{J.~E.} \bibnamefont{Lukens}},
  \bibinfo{journal}{Phys. Rev. Lett.} \textbf{\bibinfo{volume}{66}},
  \bibinfo{pages}{810} (\bibinfo{year}{1991}).

\bibitem[{\citenamefont{Leggett et~al.}(1987)\citenamefont{Leggett,
  Chakravarty, Dorsey, Fisher, Garg, and Zwerger}}]{Leggett:1987}
\bibinfo{author}{\bibfnamefont{A.~J.} \bibnamefont{Leggett}},
  \bibinfo{author}{\bibfnamefont{S.}~\bibnamefont{Chakravarty}},
  \bibinfo{author}{\bibfnamefont{A.~T.} \bibnamefont{Dorsey}},
  \bibinfo{author}{\bibfnamefont{M.~P.~A.} \bibnamefont{Fisher}},
  \bibinfo{author}{\bibfnamefont{A.}~\bibnamefont{Garg}}, \bibnamefont{and}
  \bibinfo{author}{\bibfnamefont{W.}~\bibnamefont{Zwerger}},
  \bibinfo{journal}{Rev. Mod. Phys.} \textbf{\bibinfo{volume}{59}},
  \bibinfo{pages}{1} (\bibinfo{year}{1987}).

\bibitem[{\citenamefont{{Strathearn} et~al.}(2018)\citenamefont{{Strathearn},
  {Kirton}, {Kilda}, {Keeling}, and {Lovett}}}]{Strathearn:2018}
\bibinfo{author}{\bibfnamefont{A.}~\bibnamefont{{Strathearn}}},
  \bibinfo{author}{\bibfnamefont{P.}~\bibnamefont{{Kirton}}},
  \bibinfo{author}{\bibfnamefont{D.}~\bibnamefont{{Kilda}}},
  \bibinfo{author}{\bibfnamefont{J.}~\bibnamefont{{Keeling}}},
  \bibnamefont{and} \bibinfo{author}{\bibfnamefont{B.~W.}
  \bibnamefont{{Lovett}}}, \bibinfo{journal}{Nat. Commun.}
  \textbf{\bibinfo{volume}{9}}, \bibinfo{eid}{3322} (\bibinfo{year}{2018}).

\bibitem[{\citenamefont{Anderson et~al.}(1970)\citenamefont{Anderson, Yuval,
  and Hamann}}]{Anderson:1970}
\bibinfo{author}{\bibfnamefont{P.~W.} \bibnamefont{Anderson}},
  \bibinfo{author}{\bibfnamefont{G.}~\bibnamefont{Yuval}}, \bibnamefont{and}
  \bibinfo{author}{\bibfnamefont{D.~R.} \bibnamefont{Hamann}},
  \bibinfo{journal}{Phys. Rev. B} \textbf{\bibinfo{volume}{1}},
  \bibinfo{pages}{4464} (\bibinfo{year}{1970}).

\bibitem[{\citenamefont{Magazz{\`u} et~al.}(2018)\citenamefont{Magazz{\`u},
  Forn-D{\'\i}az, Belyansky, Orgiazzi, Yurtalan, Otto, Lupascu, Wilson, and
  Grifoni}}]{Magazzu:NatCom:2018}
\bibinfo{author}{\bibfnamefont{L.}~\bibnamefont{Magazz{\`u}}},
  \bibinfo{author}{\bibfnamefont{P.}~\bibnamefont{Forn-D{\'\i}az}},
  \bibinfo{author}{\bibfnamefont{R.}~\bibnamefont{Belyansky}},
  \bibinfo{author}{\bibfnamefont{J.~L.} \bibnamefont{Orgiazzi}},
  \bibinfo{author}{\bibfnamefont{M.~A.} \bibnamefont{Yurtalan}},
  \bibinfo{author}{\bibfnamefont{M.~R.} \bibnamefont{Otto}},
  \bibinfo{author}{\bibfnamefont{A.}~\bibnamefont{Lupascu}},
  \bibinfo{author}{\bibfnamefont{C.~M.} \bibnamefont{Wilson}},
  \bibnamefont{and} \bibinfo{author}{\bibfnamefont{M.}~\bibnamefont{Grifoni}},
  \bibinfo{journal}{Nat. Commun.} \textbf{\bibinfo{volume}{9}},
  \bibinfo{pages}{1403} (\bibinfo{year}{2018}).

\bibitem[{\citenamefont{Bray and Moore}(1982)}]{Bray:1982}
\bibinfo{author}{\bibfnamefont{A.~J.} \bibnamefont{Bray}} \bibnamefont{and}
  \bibinfo{author}{\bibfnamefont{M.~A.} \bibnamefont{Moore}},
  \bibinfo{journal}{Phys. Rev. Lett.} \textbf{\bibinfo{volume}{49}},
  \bibinfo{pages}{1545} (\bibinfo{year}{1982}).

\bibitem[{\citenamefont{Chakravarty}(1982)}]{Chakravarty:1982}
\bibinfo{author}{\bibfnamefont{S.}~\bibnamefont{Chakravarty}},
  \bibinfo{journal}{Phys. Rev. Lett.} \textbf{\bibinfo{volume}{49}},
  \bibinfo{pages}{681} (\bibinfo{year}{1982}).

\bibitem[{\citenamefont{{Wallraff} et~al.}(2004)\citenamefont{{Wallraff},
  {Schuster}, {Blais}, {Frunzio}, {Huang}, {Majer}, {Kumar}, {Girvin}, and
  {Schoelkopf}}}]{Wallraff:2004}
\bibinfo{author}{\bibfnamefont{A.}~\bibnamefont{{Wallraff}}},
  \bibinfo{author}{\bibfnamefont{D.~I.} \bibnamefont{{Schuster}}},
  \bibinfo{author}{\bibfnamefont{A.}~\bibnamefont{{Blais}}},
  \bibinfo{author}{\bibfnamefont{L.}~\bibnamefont{{Frunzio}}},
  \bibinfo{author}{\bibfnamefont{R.~S.} \bibnamefont{{Huang}}},
  \bibinfo{author}{\bibfnamefont{J.}~\bibnamefont{{Majer}}},
  \bibinfo{author}{\bibfnamefont{S.}~\bibnamefont{{Kumar}}},
  \bibinfo{author}{\bibfnamefont{S.~M.} \bibnamefont{{Girvin}}},
  \bibnamefont{and} \bibinfo{author}{\bibfnamefont{R.~J.}
  \bibnamefont{{Schoelkopf}}}, \bibinfo{journal}{Nature}
  \textbf{\bibinfo{volume}{431}}, \bibinfo{pages}{162} (\bibinfo{year}{2004}).

\bibitem[{\citenamefont{Frisk~Kockum et~al.}(2019)\citenamefont{Frisk~Kockum,
  Miranowicz, De~Liberato, Savasta, and
  Nori}}]{Frisk-Kockum:NatureReviewsPhysics:2019}
\bibinfo{author}{\bibfnamefont{A.}~\bibnamefont{Frisk~Kockum}},
  \bibinfo{author}{\bibfnamefont{A.}~\bibnamefont{Miranowicz}},
  \bibinfo{author}{\bibfnamefont{S.}~\bibnamefont{De~Liberato}},
  \bibinfo{author}{\bibfnamefont{S.}~\bibnamefont{Savasta}}, \bibnamefont{and}
  \bibinfo{author}{\bibfnamefont{F.}~\bibnamefont{Nori}},
  \bibinfo{journal}{Nat. Rev. Phys.} \textbf{\bibinfo{volume}{1}},
  \bibinfo{pages}{19} (\bibinfo{year}{2019}).

\bibitem[{\citenamefont{Forn-D\'{\i}az
  et~al.}(2019)\citenamefont{Forn-D\'{\i}az, Lamata, Rico, Kono, and
  Solano}}]{FornDiaz:RevModPhys:2019}
\bibinfo{author}{\bibfnamefont{P.}~\bibnamefont{Forn-D\'{\i}az}},
  \bibinfo{author}{\bibfnamefont{L.}~\bibnamefont{Lamata}},
  \bibinfo{author}{\bibfnamefont{E.}~\bibnamefont{Rico}},
  \bibinfo{author}{\bibfnamefont{J.}~\bibnamefont{Kono}}, \bibnamefont{and}
  \bibinfo{author}{\bibfnamefont{E.}~\bibnamefont{Solano}},
  \bibinfo{journal}{Rev. Mod. Phys.} \textbf{\bibinfo{volume}{91}},
  \bibinfo{pages}{025005} (\bibinfo{year}{2019}).

\bibitem[{\citenamefont{Forn-D{\'\i}az
  et~al.}(2016)\citenamefont{Forn-D{\'\i}az, Garc{\'\i}a-Ripoll, Peropadre,
  Orgiazzi, Yurtalan, Belyansky, Wilson, and Lupascu}}]{Forn-Diaz:NatPhys:2016}
\bibinfo{author}{\bibfnamefont{P.}~\bibnamefont{Forn-D{\'\i}az}},
  \bibinfo{author}{\bibfnamefont{J.~J.} \bibnamefont{Garc{\'\i}a-Ripoll}},
  \bibinfo{author}{\bibfnamefont{B.}~\bibnamefont{Peropadre}},
  \bibinfo{author}{\bibfnamefont{J.~L.} \bibnamefont{Orgiazzi}},
  \bibinfo{author}{\bibfnamefont{M.~A.} \bibnamefont{Yurtalan}},
  \bibinfo{author}{\bibfnamefont{R.}~\bibnamefont{Belyansky}},
  \bibinfo{author}{\bibfnamefont{C.~M.} \bibnamefont{Wilson}},
  \bibnamefont{and} \bibinfo{author}{\bibfnamefont{A.}~\bibnamefont{Lupascu}},
  \bibinfo{journal}{Nat. Phys.} \textbf{\bibinfo{volume}{13}},
  \bibinfo{pages}{39 EP } (\bibinfo{year}{2016}).

\bibitem[{\citenamefont{Weiss}(2012)}]{Weiss:book}
\bibinfo{author}{\bibfnamefont{U.}~\bibnamefont{Weiss}},
  \emph{\bibinfo{title}{{Quantum Dissipative Systems; 4th ed.}}}
  (\bibinfo{publisher}{World Scientific}, \bibinfo{address}{Singapore},
  \bibinfo{year}{2012}).

\bibitem[{\citenamefont{Laird et~al.}(1991)\citenamefont{Laird, Budimir, and
  Skinner}}]{Laird:JCP:1991}
\bibinfo{author}{\bibfnamefont{B.~B.} \bibnamefont{Laird}},
  \bibinfo{author}{\bibfnamefont{J.}~\bibnamefont{Budimir}}, \bibnamefont{and}
  \bibinfo{author}{\bibfnamefont{J.~L.} \bibnamefont{Skinner}},
  \bibinfo{journal}{J. Chem. Phys.} \textbf{\bibinfo{volume}{94}},
  \bibinfo{pages}{4391} (\bibinfo{year}{1991}).

\bibitem[{\citenamefont{Reichman and Silbey}(1996)}]{Reichman:JCP:1996}
\bibinfo{author}{\bibfnamefont{D.~R.} \bibnamefont{Reichman}} \bibnamefont{and}
  \bibinfo{author}{\bibfnamefont{R.~J.} \bibnamefont{Silbey}},
  \bibinfo{journal}{J. Chem. Phys.} \textbf{\bibinfo{volume}{104}},
  \bibinfo{pages}{1506} (\bibinfo{year}{1996}).

\bibitem[{\citenamefont{Guarnieri et~al.}(2018)\citenamefont{Guarnieri,
  Kol\'a\ifmmode~\check{r}\else \v{r}\fi{}, and Filip}}]{Guarnieri:2018}
\bibinfo{author}{\bibfnamefont{G.}~\bibnamefont{Guarnieri}},
  \bibinfo{author}{\bibfnamefont{M.}~\bibnamefont{Kol\'a\ifmmode~\check{r}\else
  \v{r}\fi{}}}, \bibnamefont{and}
  \bibinfo{author}{\bibfnamefont{R.}~\bibnamefont{Filip}},
  \bibinfo{journal}{Phys. Rev. Lett.} \textbf{\bibinfo{volume}{121}},
  \bibinfo{pages}{070401} (\bibinfo{year}{2018}).

\bibitem[{\citenamefont{Zhao et~al.}(2014)\citenamefont{Zhao, Yao, Chernyak,
  and Zhao}}]{Zhao:2014}
\bibinfo{author}{\bibfnamefont{Y.}~\bibnamefont{Zhao}},
  \bibinfo{author}{\bibfnamefont{Y.}~\bibnamefont{Yao}},
  \bibinfo{author}{\bibfnamefont{V.}~\bibnamefont{Chernyak}}, \bibnamefont{and}
  \bibinfo{author}{\bibfnamefont{Y.}~\bibnamefont{Zhao}}, \bibinfo{journal}{J.
  Chem. Phys.} \textbf{\bibinfo{volume}{140}}, \bibinfo{pages}{161105}
  (\bibinfo{year}{2014}).

\bibitem[{\citenamefont{Zhou et~al.}(2015)\citenamefont{Zhou, Chen, Xu,
  Chernyak, and Zhao}}]{Zhou:PhysRev:2015}
\bibinfo{author}{\bibfnamefont{N.}~\bibnamefont{Zhou}},
  \bibinfo{author}{\bibfnamefont{L.}~\bibnamefont{Chen}},
  \bibinfo{author}{\bibfnamefont{D.}~\bibnamefont{Xu}},
  \bibinfo{author}{\bibfnamefont{V.}~\bibnamefont{Chernyak}}, \bibnamefont{and}
  \bibinfo{author}{\bibfnamefont{Y.}~\bibnamefont{Zhao}},
  \bibinfo{journal}{Phys. Rev. B} \textbf{\bibinfo{volume}{91}},
  \bibinfo{pages}{195129} (\bibinfo{year}{2015}).

\bibitem[{\citenamefont{Makri and Makarov}(1995)}]{Makri:JCP:1995}
\bibinfo{author}{\bibfnamefont{N.}~\bibnamefont{Makri}} \bibnamefont{and}
  \bibinfo{author}{\bibfnamefont{D.~E.} \bibnamefont{Makarov}},
  \bibinfo{journal}{J. Chem. Phys.} \textbf{\bibinfo{volume}{102}},
  \bibinfo{pages}{4611} (\bibinfo{year}{1995}).

\bibitem[{\citenamefont{Sim and Makri}(1997)}]{Sim:CPC:1997}
\bibinfo{author}{\bibfnamefont{E.}~\bibnamefont{Sim}} \bibnamefont{and}
  \bibinfo{author}{\bibfnamefont{N.}~\bibnamefont{Makri}},
  \bibinfo{journal}{Comput. Phys. Commun.} \textbf{\bibinfo{volume}{99}},
  \bibinfo{pages}{335} (\bibinfo{year}{1997}).

\bibitem[{\citenamefont{Sim}(2001)}]{Sim:JCP:2001}
\bibinfo{author}{\bibfnamefont{E.}~\bibnamefont{Sim}}, \bibinfo{journal}{J.
  Chem. Phys.} \textbf{\bibinfo{volume}{115}}, \bibinfo{pages}{4450}
  (\bibinfo{year}{2001}).

\bibitem[{\citenamefont{Richter and Fingerhut}(2017)}]{Richter:2017}
\bibinfo{author}{\bibfnamefont{M.}~\bibnamefont{Richter}} \bibnamefont{and}
  \bibinfo{author}{\bibfnamefont{B.~P.} \bibnamefont{Fingerhut}},
  \bibinfo{journal}{J. Chem. Phys.} \textbf{\bibinfo{volume}{146}},
  \bibinfo{pages}{214101} (\bibinfo{year}{2017}).

\bibitem[{\citenamefont{Richter and
  Fingerhut}(2019)}]{Richter:FaradayDisc:2019}
\bibinfo{author}{\bibfnamefont{M.}~\bibnamefont{Richter}} \bibnamefont{and}
  \bibinfo{author}{\bibfnamefont{B.~P.} \bibnamefont{Fingerhut}},
  \bibinfo{journal}{Faraday Discuss.} \textbf{\bibinfo{volume}{216}},
  \bibinfo{pages}{72} (\bibinfo{year}{2019}).

\bibitem[{\citenamefont{Makarov and Makri}(1994)}]{Makarov:CPL:1994}
\bibinfo{author}{\bibfnamefont{D.~E.} \bibnamefont{Makarov}} \bibnamefont{and}
  \bibinfo{author}{\bibfnamefont{N.}~\bibnamefont{Makri}},
  \bibinfo{journal}{Chem. Phys. Lett.} \textbf{\bibinfo{volume}{221}},
  \bibinfo{pages}{482} (\bibinfo{year}{1994}).

\bibitem[{\citenamefont{Braun et~al.}(2001)\citenamefont{Braun, Haake, and
  Strunz}}]{Braun:PhysRevLett:2001}
\bibinfo{author}{\bibfnamefont{D.}~\bibnamefont{Braun}},
  \bibinfo{author}{\bibfnamefont{F.}~\bibnamefont{Haake}}, \bibnamefont{and}
  \bibinfo{author}{\bibfnamefont{W.~T.} \bibnamefont{Strunz}},
  \bibinfo{journal}{Phys. Rev. Lett.} \textbf{\bibinfo{volume}{86}},
  \bibinfo{pages}{2913} (\bibinfo{year}{2001}).

\bibitem[{\citenamefont{Tuorila et~al.}(2019)\citenamefont{Tuorila,
  Stockburger, Ala-Nissila, Ankerhold, and
  M\"ott\"onen}}]{Tuorila:PhysRevResearch:2019}
\bibinfo{author}{\bibfnamefont{J.}~\bibnamefont{Tuorila}},
  \bibinfo{author}{\bibfnamefont{J.}~\bibnamefont{Stockburger}},
  \bibinfo{author}{\bibfnamefont{T.}~\bibnamefont{Ala-Nissila}},
  \bibinfo{author}{\bibfnamefont{J.}~\bibnamefont{Ankerhold}},
  \bibnamefont{and}
  \bibinfo{author}{\bibfnamefont{M.}~\bibnamefont{M\"ott\"onen}},
  \bibinfo{journal}{Phys. Rev. Research} \textbf{\bibinfo{volume}{1}},
  \bibinfo{pages}{013004} (\bibinfo{year}{2019}).

\bibitem[{\citenamefont{Grifoni et~al.}(1999)\citenamefont{Grifoni, Paladino,
  and Weiss}}]{Grifoni:1999}
\bibinfo{author}{\bibfnamefont{M.}~\bibnamefont{Grifoni}},
  \bibinfo{author}{\bibfnamefont{E.}~\bibnamefont{Paladino}}, \bibnamefont{and}
  \bibinfo{author}{\bibfnamefont{U.}~\bibnamefont{Weiss}},
  \bibinfo{journal}{Eur. Phys. J. B} \textbf{\bibinfo{volume}{10}},
  \bibinfo{pages}{719} (\bibinfo{year}{1999}).

\bibitem[{\citenamefont{L{\"u} et~al.}(2013)\citenamefont{L{\"u}, Duan, Li,
  Shenai, and Zhao}}]{Lu:JCP:2013}
\bibinfo{author}{\bibfnamefont{Z.}~\bibnamefont{L{\"u}}},
  \bibinfo{author}{\bibfnamefont{L.}~\bibnamefont{Duan}},
  \bibinfo{author}{\bibfnamefont{X.}~\bibnamefont{Li}},
  \bibinfo{author}{\bibfnamefont{P.~M.} \bibnamefont{Shenai}},
  \bibnamefont{and} \bibinfo{author}{\bibfnamefont{Y.}~\bibnamefont{Zhao}},
  \bibinfo{journal}{J. Chem. Phys.} \textbf{\bibinfo{volume}{139}},
  \bibinfo{pages}{164103} (\bibinfo{year}{2013}).

\bibitem[{\citenamefont{Romero-Rochin and
  Oppenheim}(1989)}]{Romero-Rochin:PhysicaA:1989}
\bibinfo{author}{\bibfnamefont{V.}~\bibnamefont{Romero-Rochin}}
  \bibnamefont{and}
  \bibinfo{author}{\bibfnamefont{I.}~\bibnamefont{Oppenheim}},
  \bibinfo{journal}{Physica A} \textbf{\bibinfo{volume}{155}},
  \bibinfo{pages}{52} (\bibinfo{year}{1989}).

\bibitem[{\citenamefont{Costi and Zar\'and}(1999)}]{Costi:1999}
\bibinfo{author}{\bibfnamefont{T.~A.} \bibnamefont{Costi}} \bibnamefont{and}
  \bibinfo{author}{\bibfnamefont{G.}~\bibnamefont{Zar\'and}},
  \bibinfo{journal}{Phys. Rev. B} \textbf{\bibinfo{volume}{59}},
  \bibinfo{pages}{12398} (\bibinfo{year}{1999}).

\bibitem[{\citenamefont{Ruokola and Ojanen}(2011)}]{Ruakola:PhysRevB:2011}
\bibinfo{author}{\bibfnamefont{T.}~\bibnamefont{Ruokola}} \bibnamefont{and}
  \bibinfo{author}{\bibfnamefont{T.}~\bibnamefont{Ojanen}},
  \bibinfo{journal}{Phys. Rev. B} \textbf{\bibinfo{volume}{83}},
  \bibinfo{pages}{045417} (\bibinfo{year}{2011}).

\bibitem[{\citenamefont{Kast and Ankerhold}(2013)}]{Kast:2013}
\bibinfo{author}{\bibfnamefont{D.}~\bibnamefont{Kast}} \bibnamefont{and}
  \bibinfo{author}{\bibfnamefont{J.}~\bibnamefont{Ankerhold}},
  \bibinfo{journal}{Phys. Rev. Lett.} \textbf{\bibinfo{volume}{110}},
  \bibinfo{pages}{010402} (\bibinfo{year}{2013}).

\bibitem[{\citenamefont{Garg et~al.}(1985)\citenamefont{Garg, Onuchic, and
  Ambegaokar}}]{Garg:JCP:1985}
\bibinfo{author}{\bibfnamefont{A.}~\bibnamefont{Garg}},
  \bibinfo{author}{\bibfnamefont{J.}~\bibnamefont{Onuchic}}, \bibnamefont{and}
  \bibinfo{author}{\bibfnamefont{V.}~\bibnamefont{Ambegaokar}},
  \bibinfo{journal}{J. Chem. Phys.} \textbf{\bibinfo{volume}{83}},
  \bibinfo{pages}{4491} (\bibinfo{year}{1985}).

\bibitem[{\citenamefont{Lambert et~al.}(2019)\citenamefont{Lambert, Ahmed,
  Cirio, and Nori}}]{Lambert:NatCom:2019}
\bibinfo{author}{\bibfnamefont{N.}~\bibnamefont{Lambert}},
  \bibinfo{author}{\bibfnamefont{S.}~\bibnamefont{Ahmed}},
  \bibinfo{author}{\bibfnamefont{M.}~\bibnamefont{Cirio}}, \bibnamefont{and}
  \bibinfo{author}{\bibfnamefont{F.}~\bibnamefont{Nori}},
  \bibinfo{journal}{Nat. Commun.} \textbf{\bibinfo{volume}{10}},
  \bibinfo{pages}{3721} (\bibinfo{year}{2019}).

\bibitem[{\citenamefont{Correa et~al.}(2019)\citenamefont{Correa, Xu, Morris,
  and Adesso}}]{Correa:JCP:2019}
\bibinfo{author}{\bibfnamefont{L.~A.} \bibnamefont{Correa}},
  \bibinfo{author}{\bibfnamefont{B.}~\bibnamefont{Xu}},
  \bibinfo{author}{\bibfnamefont{B.}~\bibnamefont{Morris}}, \bibnamefont{and}
  \bibinfo{author}{\bibfnamefont{G.}~\bibnamefont{Adesso}},
  \bibinfo{journal}{J. Chem. Phys.} \textbf{\bibinfo{volume}{151}},
  \bibinfo{pages}{094107} (\bibinfo{year}{2019}).

\bibitem[{\citenamefont{Chernyak and Mukamel}(1996)}]{Chernyak:1996}
\bibinfo{author}{\bibfnamefont{V.}~\bibnamefont{Chernyak}} \bibnamefont{and}
  \bibinfo{author}{\bibfnamefont{S.}~\bibnamefont{Mukamel}},
  \bibinfo{journal}{J. Chem. Phys.} \textbf{\bibinfo{volume}{105}},
  \bibinfo{pages}{4565} (\bibinfo{year}{1996}).

\bibitem[{\citenamefont{Mavros et~al.}(2016)\citenamefont{Mavros, Hait, and
  Van~Voorhis}}]{Mavros:2016}
\bibinfo{author}{\bibfnamefont{M.~G.} \bibnamefont{Mavros}},
  \bibinfo{author}{\bibfnamefont{D.}~\bibnamefont{Hait}}, \bibnamefont{and}
  \bibinfo{author}{\bibfnamefont{T.}~\bibnamefont{Van~Voorhis}},
  \bibinfo{journal}{J. Chem. Phys.} \textbf{\bibinfo{volume}{145}},
  \bibinfo{pages}{214105} (\bibinfo{year}{2016}).

\bibitem[{\citenamefont{Domcke et~al.}(2004)\citenamefont{Domcke, Yarkony, and
  K{\"o}ppel}}]{Domcke:Book}
\bibinfo{author}{\bibfnamefont{W.}~\bibnamefont{Domcke}},
  \bibinfo{author}{\bibfnamefont{D.~R.} \bibnamefont{Yarkony}},
  \bibnamefont{and}
  \bibinfo{author}{\bibfnamefont{H.}~\bibnamefont{K{\"o}ppel}},
  \emph{\bibinfo{title}{Conical Intersections}}, Advanced Series in Physical
  Chemistry (\bibinfo{publisher}{World Scientific}, \bibinfo{year}{2004}).

\bibitem[{\citenamefont{Tamura et~al.}(2012)\citenamefont{Tamura, Martinazzo,
  Ruckenbauer, and Burghardt}}]{Tamura:JCP:2012}
\bibinfo{author}{\bibfnamefont{H.}~\bibnamefont{Tamura}},
  \bibinfo{author}{\bibfnamefont{R.}~\bibnamefont{Martinazzo}},
  \bibinfo{author}{\bibfnamefont{M.}~\bibnamefont{Ruckenbauer}},
  \bibnamefont{and}
  \bibinfo{author}{\bibfnamefont{I.}~\bibnamefont{Burghardt}},
  \bibinfo{journal}{J. Chem. Phys.} \textbf{\bibinfo{volume}{137}},
  \bibinfo{pages}{22A540} (\bibinfo{year}{2012}).

\bibitem[{\citenamefont{Christensson et~al.}(2012)\citenamefont{Christensson,
  Kauffmann, Pullerits, and Man{\v c}al}}]{Christensson:2012aa}
\bibinfo{author}{\bibfnamefont{N.}~\bibnamefont{Christensson}},
  \bibinfo{author}{\bibfnamefont{H.~F.} \bibnamefont{Kauffmann}},
  \bibinfo{author}{\bibfnamefont{T.}~\bibnamefont{Pullerits}},
  \bibnamefont{and} \bibinfo{author}{\bibfnamefont{T.}~\bibnamefont{Man{\v
  c}al}}, \bibinfo{journal}{J. Phys. Chem. B} \textbf{\bibinfo{volume}{116}},
  \bibinfo{pages}{7449} (\bibinfo{year}{2012}).

\bibitem[{\citenamefont{Schnedermann et~al.}(2019)\citenamefont{Schnedermann,
  Alvertis, Wende, Lukman, Feng, Schr{\"o}der, Turban, Wu, Hine, Greenham
  et~al.}}]{Schnedermann:NatComm:2019}
\bibinfo{author}{\bibfnamefont{C.}~\bibnamefont{Schnedermann}},
  \bibinfo{author}{\bibfnamefont{A.~M.} \bibnamefont{Alvertis}},
  \bibinfo{author}{\bibfnamefont{T.}~\bibnamefont{Wende}},
  \bibinfo{author}{\bibfnamefont{S.}~\bibnamefont{Lukman}},
  \bibinfo{author}{\bibfnamefont{J.}~\bibnamefont{Feng}},
  \bibinfo{author}{\bibfnamefont{F.~A. Y.~N.} \bibnamefont{Schr{\"o}der}},
  \bibinfo{author}{\bibfnamefont{D.~H.~P.} \bibnamefont{Turban}},
  \bibinfo{author}{\bibfnamefont{J.}~\bibnamefont{Wu}},
  \bibinfo{author}{\bibfnamefont{N.~D.~M.} \bibnamefont{Hine}},
  \bibinfo{author}{\bibfnamefont{N.~C.} \bibnamefont{Greenham}},
  \bibnamefont{et~al.}, \bibinfo{journal}{Nat. Commun.}
  \textbf{\bibinfo{volume}{10}}, \bibinfo{pages}{4207} (\bibinfo{year}{2019}).

\bibitem[{\citenamefont{Blais et~al.}(2004)\citenamefont{Blais, Huang,
  Wallraff, Girvin, and Schoelkopf}}]{Blais:2004}
\bibinfo{author}{\bibfnamefont{A.}~\bibnamefont{Blais}},
  \bibinfo{author}{\bibfnamefont{R.-S.} \bibnamefont{Huang}},
  \bibinfo{author}{\bibfnamefont{A.}~\bibnamefont{Wallraff}},
  \bibinfo{author}{\bibfnamefont{S.~M.} \bibnamefont{Girvin}},
  \bibnamefont{and} \bibinfo{author}{\bibfnamefont{R.~J.}
  \bibnamefont{Schoelkopf}}, \bibinfo{journal}{Phys. Rev. A}
  \textbf{\bibinfo{volume}{69}}, \bibinfo{pages}{062320}
  (\bibinfo{year}{2004}).

\bibitem[{\citenamefont{{Florens} et~al.}(2010)\citenamefont{{Florens},
  {Venturelli}, and {Narayanan}}}]{Florens:2010}
\bibinfo{author}{\bibfnamefont{S.}~\bibnamefont{{Florens}}},
  \bibinfo{author}{\bibfnamefont{D.}~\bibnamefont{{Venturelli}}},
  \bibnamefont{and}
  \bibinfo{author}{\bibfnamefont{R.}~\bibnamefont{{Narayanan}}},
  \emph{\bibinfo{title}{Quantum Phase Transition in the Spin Boson Model}}
  (\bibinfo{publisher}{Springer}, \bibinfo{address}{Berlin, Heidelberg},
  \bibinfo{year}{2010}), pp. \bibinfo{pages}{145--162}.

\bibitem[{\citenamefont{Hamm and Stock}(2012)}]{Hamm:PhysRevLett:2012}
\bibinfo{author}{\bibfnamefont{P.}~\bibnamefont{Hamm}} \bibnamefont{and}
  \bibinfo{author}{\bibfnamefont{G.}~\bibnamefont{Stock}},
  \bibinfo{journal}{Phys. Rev. Lett.} \textbf{\bibinfo{volume}{109}},
  \bibinfo{pages}{173201} (\bibinfo{year}{2012}).

\bibitem[{\citenamefont{Eckel et~al.}(2009)\citenamefont{Eckel, Reina, and
  Thorwart}}]{Eckel:NJP:2009}
\bibinfo{author}{\bibfnamefont{J.}~\bibnamefont{Eckel}},
  \bibinfo{author}{\bibfnamefont{J.~H.} \bibnamefont{Reina}}, \bibnamefont{and}
  \bibinfo{author}{\bibfnamefont{M.}~\bibnamefont{Thorwart}},
  \bibinfo{journal}{New J. Phys.} \textbf{\bibinfo{volume}{11}},
  \bibinfo{pages}{085001} (\bibinfo{year}{2009}).

\end{thebibliography}


\end{document}